\documentstyle[11pt,aussois2003,twoside,epsf]{article}
\def\edcomment#1{\iffalse\marginpar{\raggedright\sl#1\/}\else\relax\fi}
\reversemarginpar
\begin{document}
%
\title{Reconstructing images of gravitational lenses with regularizing algorithms}
%
\author{E. Koptelova}
\affil{Sternberg State Astronomical Institute, Moscow State
University, Universitetsky Prospect, 13, 119992, Moscow, Russia}

\author{E. Shimanovskaya$^{1}$, B. Artamonov$^{2}$, V.Belokurov$^{3}$,
M. Sazhin$^{2}$, A.~Yagola$^{1}$} \affil {$^{1}$Department of
Mathematics, Faculty of Physics, Moscow State University,
Vorobyevy Gory, 119992, Moscow, Russia} \affil {$^{2}$Sternberg
State Astronomical Institute, Moscow State University,
Universitetsky Prospect, 13, 119992, Moscow, Russia} \affil
{$^{3}$Theoretical Physics, Oxford University, 1 Keble Road, OX1
3NP, Oxford, UK}

\label{page:first}
\begin{abstract}
This note addresses possible applications of the Tikhonov
regularization to image reconstruction of gravitational lens
systems. Several modifications of the regularization algorithm
are discussed. Our illustrative example is the close quadruple
gravitational lens QSO 2237+0305 (Einstein Cross). The restored
image of the lens is decomposed into two parts -- the quasar
components and the background galaxy.
\end{abstract}
\section{Introduction}
Over the last decade gravitational lensing has become a powerful
tool to probe the Universe. The number of lenses discovered so
far is just below one hundred and is growing quickly.
Gravitational lensing comes in a variety of shapes and sizes. One
of the most interesting examples of this phenomenon is systems of
close multiple quasar images. For a typical lens galaxy the image
separation is of order of an arc second which happens to be the
resolution of some of the best ground-based telscopes. Hence,
what the observer will see is the source copies blurred and
overlapped and possibly superimposed on the lens galaxy. To use
the "nature telescope" one needs to restore the light
distribution preferably splitting it into parts belonging to the
"background" and to the quasar images. The problem addressed in
this note is the problem of the correct reconstruction of complex
gravitational lens images.

\section{Algorithmic aspects of the image reconstruction}
There is a number of image reconstruction algorithms available for
astronomers at present. Let us briefly review some of them. For
example, in the CLEAN method one proceeds by subtracting the
quasar components from the image in order to obtain underlying
galaxy. The process is repeated iteratively until the termination
criterion is satisfied, which usually requires visual inspection
of the gray-scale distribution of residuals. Another wide-spread
method is the Maximum Entropy deconvolution. Here the solution
not only fits the noisy data but also maximizes so-called
'entropy'. Essentially, the entropy term is a measure of the
number of different features in the image. In some cases this
method can lead to a significant improvement in quality, however
the very definition of the entropy forces the resulting image to
be smooth, which sometimes can be considered as a drawback. In
both the CLEAN and the MEM algorithms some kind of a 'default'
image has to be introduced. The approach developed by Magain,
Courbin and Sohy (MCS) requires only the PSF of the final
(restored) image to be set by hand. The MCS method proved to work
well for the data on which the previous algorithms had failed to
give the optimal results. However, problems might appear when
the MCS algorithm is faced with the image with irregular or
varying over the field point spread function (PSF).

\subsection{Deconvolution basics}

The mathematical model of the image corruption due to the
finite resolving power of the telescope and atmospheric
perturbation is a Fredholm type integral equation of the first
kind with a space-invariant kernel, or a convolution integral
equation:
\begin{equation}
\label{fredholm} A[z](x,y)=\int\!\!\!\int\limits_{\!\!B}
t(x-\xi,y-\eta)z(\xi,\eta)\,d\xi d\eta=u (x,y)
\end{equation}
Here $u (x,y)$ represents the observed light distribution,
$t(x,y)$ is the point spread function determined from
observations of the individual star at the periphery of the
frame, $z(x,y)$ represents the object, or sought solution, $B$ is
the frame area, $B=[0,L]\times[0,L]$. The convolution operator
$A$ is the linear operator which acts from some Hilbert space $Z$
to the Hilbert space of second-power-integrable functions $L_{2}$.
The goal is to find an approximate solution having at our
disposal the noisy data $u_\delta$ and the estimate
of the noise level $\delta$: $\|u-u_\delta \|_{L_2}\leq\delta$.
This inverse problem belongs to the general class of ill-posed
problems in the sense of Hadamar.

There is a well-developed mathematical approach to solve such
problems based on the idea of the regularizing algorithm. Since
the publication of the fundamental work by A. Tikhonov (1963)
this method has been extensively studied and widely adopted in many
fields of science.

\subsection{A regularizing algorithm}

To obtain stable and physically valid results in image reconstruction
the regularization ought to be based on {\it a priori} knowledge of
the properties of the admissible solution.

The regularization implies construction of the algorithm that controls
the trade-off between the assumptions about smoothness and the
structure of the sought solution and its consistency with the
data. The key concept of the algorithm is a smoothing function:
\begin{equation}
\label{regular}
M^\alpha[z]=\|A[z]-u_{\delta}\|^{2}_{L_2}+\alpha\cdot\Omega[z]
\end{equation}
Here the first term represents the squared discrepancy, $\alpha$ is
the regularization parameter, $\Omega[z]$ is a stabilizer function.
Let $z^{\alpha}$ be the extremum of the function $M^\alpha[z]$ on $Z$,
i.e. $z^{\alpha}$ is the solution of the minimization problem for
$M^\alpha[z]$ on the chosen set of functions (possibly with some
constraints).

The choice of the regularization parameter $\alpha$ is crucial for
solving ill-posed problems. Generally, it should depend on the input
data, the errors, and the method of approximation of the initial
problem. One of the way to co-ordinate the regularization parameter
with the error of the input information is the discrepancy principle:
\begin{equation}
\label{discprinc} \|A[z^{\alpha}]-u_{\delta}\|_{L_2}\simeq\delta.
\end{equation}

Providing the regularization parameter $\alpha$ is chosen according to
this rule, $z^{\alpha}$ can be considered as an approximate solution
which tends to the exact solution in the context of the norm of the
chosen set of functions as the error level of input data tends to
zero.


A prior knowledge about the smoothness of the unknown solution is
embedded in the regularizing algorithm through the appropriate choice
of the stabilizer function. In most cases it is the squared norm of
the solution on some set of functions: $\Omega[z]=\|z\|^2_{Z}$. The
choice of the stabilizer affects the order of convergence of
approximate solutions. Thus, if it is assumed that the unknown
solution belongs to the class of second-power-integrable functions,
$Z\equiv L_2$, the stabilizer can be chosen as follows:
\begin{equation}
\label{l2stab} \Omega[z]=\|z\|^2_{L_2}\equiv
\int\!\!\!\int\limits_{\!\!B} z^2 dx dy,
\end{equation}
Regularization with this stabilizer function guarantees that
approximate solutions converge to the exact solution in the
context of the norm $L_2$, i.e. converges in mean
(zero-order convergence):\\
$$ \|z^\alpha-z\|_{L_2}\rightarrow 0 \quad as \quad \delta\rightarrow
0.
$$
If {\it a priori} information about sought solution allows to
assume higher smoothness of $z$ and choose $Z\equiv W_{22}$, where
$W_{22}$ is a set of $L_2$-functions which have generalized
derivatives of the second order those are second-power-integrable, the
stabilizer can be written in the following form:
\begin{equation}
\label{w22stab} \Omega[z]=\|z\|^2_{W_{22 }}\equiv
\int\!\!\!\int\limits_{\!\!B} \left\{ z^2+\left(\frac{\partial^2
z}{\partial x^2}\right)^2+2\left(\frac{\partial^2 z}{\partial x
\partial y}\right)^2+\left(\frac{\partial^2 z}{\partial
y^2}\right)^2 \right\} dx dy.
\end{equation}
When selecting $\alpha$ in accordance with the discrepancy principle,
approximate solutions $z^\alpha$ tend to the exact solution of the
problem as $\delta$ tends to zero in the context of the $W_{22}$
norm:
$$
\|z^\alpha-z\|_{W_{22}}\rightarrow 0 \quad as \quad
\delta\rightarrow 0.
$$
According to Sobolev's embedding theorem, $W_{22}[B]$ is embedded in
$C[B]$~-- the set of continuous functions on $B$. Thus, the
convergence in the context of the $W_{22}$-norm means the convergence
in the context of the norm of $C[B]$, i.e. regularized solutions
converge to the exact solution uniformly:
$$\max_{(x,y)\in B}|z^\alpha(x,y)-z(x,y)| \rightarrow 0
\quad as \quad \delta \rightarrow 0$$

The degree of smoothness of the sought solution is not the only {\it a
priori} information that can be considered and included into the
regularizing algorithm. Various assumptions about the structure of the
object under study can also be taken into account. Images of close
quadruple gravitational lens systems consist of multiple overlapped
quasar images superimposed on a background galaxy. So the image can be
decomposed into two constituent parts - the sum of K
$\delta$-functions and smooth background (galaxy):
\begin{equation}
\label{splitting}
 z(x,y)=\sum_{k=1}^{K}a_k
\delta(x-b_k,y-c_k)+g(x,y),
\end{equation}
where $K$ is the number of point sources with coordinates $(b_k,c_k)$
and intensities $a_k$ in the frame; $g(x,y)$ is the solution's
component corresponding to a galaxy; $\delta$ represents Dirac
function.

Rapid intensity variations in the observed data caused by the bright
nucleus of the galaxy can be processed by incorporation of an
additional $\delta$-function for the nucleus or by selecting the
function describing the light distribution of the background galaxy
from the appropriate set of functions, i.e. bounded total variation
(TV) functions.

An approach of piecewise uniform regularization based on TV class of
functions was suggested by A. Leonov (1999). Let us consider an
arbitrary grid $S_{N_1 N_2}$ introduced on $B$ and define the total
variation for a function $z$ on $B$ as follows:
$$
\begin{array}{c}
V(z,B)= \sup \limits_{S_{N_1 N_2}} (\sum
\limits_{m=1}^{N_1-1}|z_{m+1,1}-z_{m,1}|+ \sum\limits_{n=1}^{N_2
-1}|z_{1,n+1}-z_{1,n}|+\\\\
 +\sum\limits_{m=1}^{N_1 -1}\sum\limits_{n=1}^{N_2 -1}|z_{m+1,n+1}-z_{m+1,n}-z_{m,n+1}+z_{m,n}|
 ,  \forall S_{N_1 N_2})
\end{array}
$$
The function for which the total variation is a finite quantity is
called bounded total variation function. It is continious nearly
everywhere with the exception, possibly, of the points of
discontinuity positioned on the countable set of gridlines. The
regularization algorithm with the proper choice of the regularization
parameter and the stabilizer function
\begin{equation}
\Omega[z]\equiv\|z\|_{\nu [B]}=|z(0,0)|+V(z,B)
\end{equation}
provides piecewise uniform convergence of approximate solutions.

Additionally, one can penalize the unknown solution for being
drastically different from the certain analytical model and construct
the stabilizer in the following form:
\begin{equation}
\label{stab}
 \Omega[z]=\|g-g_{model}\|^2_G
\end{equation}
In this work we assume that the light distribution in the central
region of the galaxy is well-modeled by generalized de Vaucouleurs
profile (Sersic's model):
\begin{equation}
\label{sersic}
 g_{model}(r)=I_0\exp^{-b_n(\frac{r}{r_e})^{\frac{1}{n}}}
\end{equation}
where $b_n=2n-0.324$ for $1\leq n \leq 4$.


\section{Numerical results}

\subsection{Observations}

The observations of QSO 2237+0305 were carried out during August,
September and October of 2002 on the 1.5-m telescope AZT-22 at the
Maidanak Observatory in Uzbekistan. The CCD camera with the gain of
$g=1.2 e^{-}ADU^{-1}$, the readout noise of $RON=10e^{-}$ and the
pixel scale of $0.12 \arcsec pixel^{-1}$ was used. In this work one of
R-band frames obtained on the 29th of August 2002 was taken. The best
quality of the image corresponds to the point source with the
FWHM=0.75\arcsec. Preprocessing of the data including bias-level
subtraction, flat-field division, sky subtraction and cosmic ray
removal was done with the standard routines in Munich Image Data
Analysis System (MIDAS) environment. The set of data taken over the
night was averaged and the constant background over the frame was
subtracted to prepare image of Einstein Cross for the further
treatment. Then the subframe of 64 by 64 pixels with the image of
Einstein Cross centered on nucleus of the galaxy 2237+0305 was
extracted. The total noise over the frame was calculated as follows:
\begin {equation}
\delta=\sqrt{\sum_{i,j}(\frac{cross_{ij}}{Ng}+\frac{RON^{2}}{N})}
\end {equation}
where N=4 is the number of the frame of one observation night;
cross$_{ij}$ - intensity of ij-pixel. For the subframe the total
noise is about 400 ADU.

\subsection{PSF processing}

In the Einstein Cross data, the quasar images lie so close to each
other that the PSF from each image overlap with the PSFs of its
neighbours. But for normal atmospheric conditions it is vital to have
a clean isolated PSF for the data frame that we want to process. The
only way then is to derive the PSF from the bright star labelled
$\alpha$ in the frame. To create the PSF the star was extracted from
the frame. It has to be noted that the shape of the PSF can be very
complicated. This includes the sum of the imaging properties of the
telescope and the effects of atmospheric turbulence on the
signal. There is a number of methods to construct a PSF. In this work
both numerical PSF and Gaussian PSF were tested. The use of a
numerical PSF gave the best fit to the quasar components. The wings of
the PSF light distribution were slightly smoothed by the median filter
with the window size compared to the FWHM. To find the parameters of
Gaussian PSF it is necessary to solve the inverse problem. However,
the approximation of the PSF by the Gauss profile is not a good
representation of the real light distribution. If $\kappa$ is the
ratio of total intensity of the model and the total intensity of the
observed star, the value of summarised intensity losses are derived as
follows:
\begin {equation}
y=(1-\kappa)100\%
\end {equation}
In the case of the Gaussian PSF the losses amount to 4.8\% and in the
case of numerical PSF slightly smoothed by filter the losses do
not exceed 3.9\%.
\begin{figure}
\plotfiddle{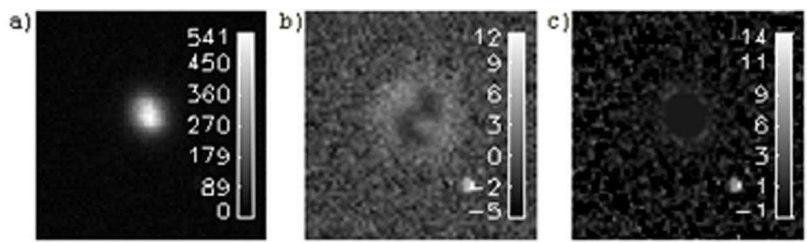}{2.5cm}{0}{80}{80}{-200}{-10}
\caption{PSF fitting: a) numerical PSF (star from frame); b)
residuals $(Gaussian PSF-numerical PSF)/\sqrt{Gaussian PSF}$; c)
residuals $(filtered PSF-numerical PSF)/\sqrt{filtered PSF}$}
\end{figure}

\subsection{Image reconstruction for QSO 2237+0305}

Working with CCD-frames one has a pixel grid naturally introduced in
the image frame and deals with discrete functions in (1). So it is
necessary to consider the finite difference approximation of the
smoothing function:
\begin{equation}
M^{\alpha}[a_k,g_{ij}]=\sum_{p,q=1}^{N}\frac{1}{\sigma_{pq}^2}\left\{\sum_{i,j=1}^{N}\left\{t_{p-i,q-j}\left(\sum_{k=1}^{4}a_k\delta_{i-b_k,j-c_k}+g_{ij}\right)\right\}-u_{pq}\right\}^2+\alpha\Omega[g_{ij}]
\end{equation}
Here the variables are the coordinates and the intensities of the
quasar components and the pixel values of the background galaxy.

\subsubsection{Image reconstruction on $L_2$}
The regularization on $L_2$ set of functions represents the simplest
sort of Tikhonov's method. Written for the pixel grid the stabilizer:
\begin{equation}
\Omega[\hat{g}_{ij}]=\sum\limits_{i,j=1}^{N}\hat{g}_{ij}^2
\end{equation}
Here $\hat{g}=g-g_{sersic}$. The parameters of the Sersic model were
obtained at preliminary stage using the least-squares method. For the
minimization of the smoothing function (12) the conjugate-gradient
method was used. Figure 2 shows the results of the image
reconstruction. The map of residuals was calculated as follows:
$(model_{ij}-cross_{ij})/{\sigma_{ij}}$.  Figure 3 shows the
astrometry results.

\begin{figure}
\label{l2fig}
\plotfiddle{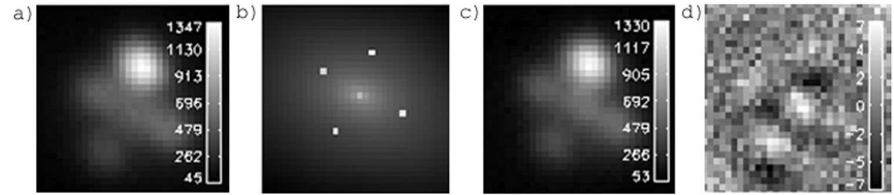}{2.5cm}{0}{90}{90}{-200}{-15}
\vspace{5pt} \caption{Reconstruction on $L_2$: a) Einstein Cross
data; b) solution on $L_2$ (logarithmic scale); c) solution
convolved with PSF; d) residuals}
\end{figure}

\begin{figure}
\plotfiddle{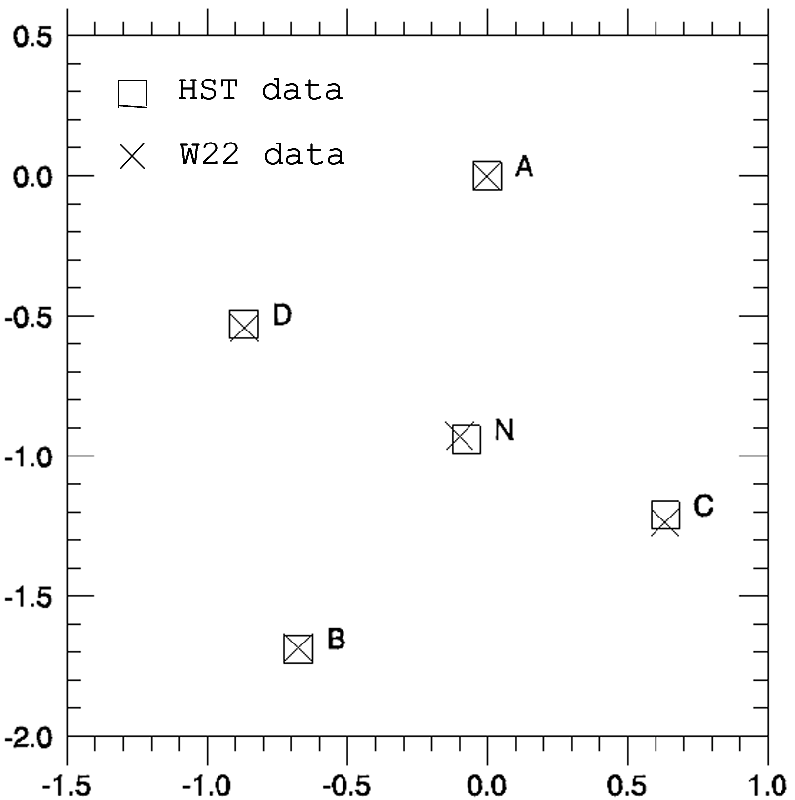}{6.5cm}{0}{90}{90}{-190}{-15} \vspace{5pt}
\caption{Astrometry for $L_2$-reconstruction}
\end{figure}

\subsubsection{Image reconstruction on $W_{22}$}
The discrete expression for the stabilizer is:
$$
\begin{array}{c}
\Omega[\hat{g}_{ij}]=\sum\limits_{i,j=1}^{N}\hat{g}_{ij}^2+
\sum\limits_{i=2}^{N-1}\sum\limits_{j=1}^{N}(\hat{g}_{i+1,j}-2\hat{g}_{i,j}+\hat{g}_{i-1,j})^2+\\\\
\sum\limits_{i=1}^{N-1}\sum\limits_{j=1}^{N-1}2(\hat{g}_{i+1,j+1}-\hat{g}_{i+1,j}-\hat{g}_{i,j+1}+\hat{g}_{i,j})^2+\sum\limits_{i=1}^{N}\sum\limits_{j=2}^{N-1}(\hat{g}_{i,j+1}-2\hat{g}_{i,j}+\hat{g}_{i,j-1})^2
\end{array}
$$
Figure 4 shows the results of the image reconstruction with the
regularizing algorithm based on the $W_{22}$ set of functions.
\begin{figure}
\label{W22fig}
\plotfiddle{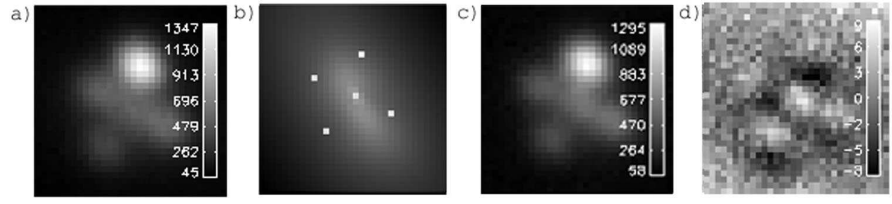}{2cm}{0}{90}{90}{-200}{-15}
\caption{Reconstruction on $W_{22}$: a) the Einstein Cross data;
b) solution on $W_{22}$ (logarithmic scale); c) solution convolved
with PSF; d) residuals}
\end{figure}

\begin{figure}
\plotfiddle{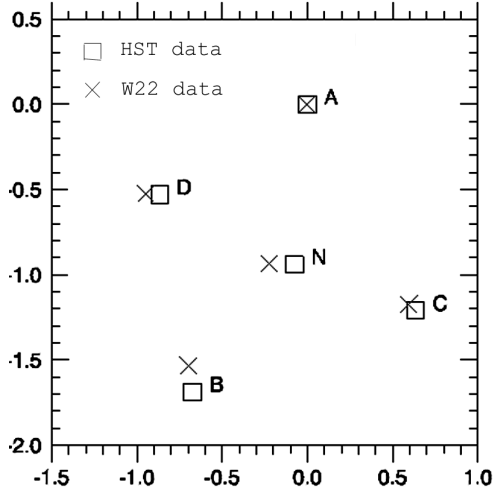}{5.5cm}{0}{80}{80}{-180}{-15}
\vspace{5pt} \caption{Astrometry for $W_{22}$-reconstruction}
\end{figure}

\subsubsection{Image reconstruction on $\nu [B]$}

The regularization technique with the use of bounded total variation
functions is applicable if there are reasons to assume that the
unknown solution has discontinuities or rapid variations. The discrete
expression for stabilizer reads:
$$
\Omega[z]=\sum_{i,j=1}^{N-1}f_\epsilon(g_{i+1,j+1}-g_{i+1,j}-g_{i,j+1}+g_{i,j})
$$
where
$f_\epsilon(t)=\sqrt{t^2+\left(\frac{\epsilon}{N^2}\right)}$.
The results of the image restoration are presented on Figure 6.
\begin{figure}
\label{tvfig} \plotfiddle{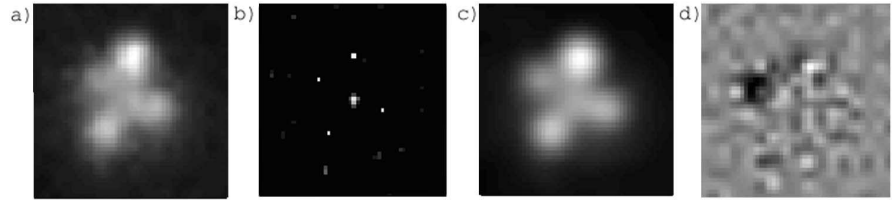}{2.5cm}{0}{90}{90}{-200}{-15}
\vspace{5pt} \caption{Reconstruction on the $\nu [B]$: a) Einstein
Cross data; b) solution on $\nu [B]$; c) solution convolved with PSF;
d) residuals}
\end{figure}

\section{Conclusions}
The application of the Tikhonov regularization to the image
reconstruction of gravitational lens systems seems to be an effective
way of tackling some of the common problems shared by conventional
deconvolution algorithms. We have considered several modifications of
the regularization method. The illustrative example of the Einstein
Cross lens system was used to compare different versions of the
algorithm. Although the results look quite encouraging the further
work needs to be done to select the appropriate stabilizer function
and to produce the pipe-line version of the software.

\acknowledgements The work was partially supported by the Russian
Foundation for Basic Research (grants 02-01-00044 and 01-02-16800).

\label{page:last}
\end{document}